\documentclass[twocolumn,pre,superscriptaddress]{revtex4}
\usepackage{graphicx,float}
\usepackage{latexsym}
\usepackage{epstopdf}
\usepackage{amssymb,amsmath}
\usepackage{color}
\usepackage{multirow}
\usepackage{hyperref}
\usepackage{color,soul}
\makeatletter
\let\Hy@linktoc\Hy@linktoc@none
\makeatother

\usepackage{verbatim}
\usepackage{bm}

\begin{document}

\title{Controlling Anderson localization in disordered heterostrctures with L\'evy-type distribution}
\author{Abbas Ghasempour Ardakani}\email{aghasempour@shirazu.ac.ir}
\author{Mohsen Ghasemi Nezhadhaghighi}
\affiliation{Department of Physics, College of Science, Shiraz University, Shiraz 71454, Iran}

\begin{abstract}

In this paper, we propose a disordered heterostructure in which the 
distribution of refractive index of one of its constituents follows a 
L\'evy-type distribution characterized by the exponent $\alpha$. 
For the normal and oblique 
incidences, the effect of $\alpha$ variation on the localization length is 
investigated in different frequency ranges. As a result, the controllability of 
Anderson localization can be achieved by changing the exponent $\alpha$ in the 
disordered structure having heavy tailed distribution.                

\end{abstract}

\maketitle

\section{Introduction}\label{sec1}

In the past few decades, localization of waves in disordered structures
known as Anderson localization has attracted growing attention, 
in various fields of physics. 
Starting from early studies of the localization of 
electronic wave functions in disordered
crystals \cite{anderson58}, it is believed 
that the localized states appear in a wide variety of classical
and quantum materials. 
The possible occurrence of
Anderson localization for electrons in disordered solids \cite{Evers2008},
 ultrasound and acoustic waves \cite{Strybulevych}, 
 the transport of light \cite{Wiersma1997,Maret2006}, microwave \cite{Dalichaouch}, 
 matter waves \cite{Billy,Roati} and cold atoms \cite{Kondov} are just a 
 few examples among many others. 
 
 All these disordered systems are 
 typically composed of regions
with very different kinds of the spatial distribution of disorder.
The question is what one would expect for
 effects of the distribution of disorder on controlling the localization 
 properties. This interesting question has been basis of numerous theoretical and
experimental studies \cite{Moura,Izrailev,Iomin,Falceto,Costa,Fernandez-Marin2012,Fernandez-Marin2013,Fernandez-Marin2014,zakeri}.

More recently, some engineered tunable random mediums have been assembled
 in the lab that follows the L\'evy type distribution
  \cite{Fernandez-Marin2012,Fernandez-Marin2014,zakeri}. 
L\'evy processes have a power-law distribution $p(x)\sim x^{-(1+\alpha)}$,
 where $\alpha$ is the so called L\'evy exponent \cite{Uchaikin}. 
Power law distributions have been appeared in different physical 
phenomena such as spectral
fluctuation in random lasers \cite{Lepro2007,Sharma2006}, 
superdiffusive transport of light 
across glass microspheres whose
diameters have levy distribution \cite{Barthelemy}, 
quantum coherent transport of electrons in one-
dimensional ($1$D) disordered wires with L\'evy-type distribution \cite{Falceto}. 

Recently, Fernandez-Marin \textit{et al.}
have investigated numerically how transmission of electromagnetic waves 
varies with the system size in a $1$D
random system in which the layer thicknesses follow a L\'evy-type 
distribution with the exponent $\alpha$ \cite{Fernandez-Marin2013}.
 They demonstrated that $\langle -\ln T \rangle \propto L^\alpha$ for
  $0<\alpha <1$ , whereas $\langle -\ln T \rangle$
is proportional to the system size $L$ for $1<\alpha <2$ \cite{Fernandez-Marin2013}.
Furthermore, Fernandez-Marın \textit{et al.} have experimentally studied the 
microwave electromagnetic wave
transmission through a waveguide composed of dielectric slabs where 
spacing between them follows
a distribution with a power-law tail ($\alpha$-stable L\'evy distribution) \cite{Fernandez-Marin2014}. 
They observed an anomalous localization for the case $0<\alpha<1$ 
in which transmission $\langle T \rangle$ decays with length of 
waveguide as $L^{-\alpha}$
while in the case of standard localization $\langle T \rangle \propto \exp (-L/\xi)$,
 $\xi$ being the localization length \cite{Fernandez-Marin2014}.

{More recently, Zakeri \textit{et al.} have investigated Anderson localization
 of the classical lattice waves in a chain with mass impurities distributed
 randomly through a power-law relation $s^{-(1+\alpha)}$
 where $s$ is the distance between two successive Impurities} \cite{zakeri}.
{ For this 1D harmonic disordered lattice of $n$-sites with random masses $m_n$, 
they have indicated that in the small frequencies for $1<\alpha<2$,
 the localization length behaves as $\xi (\omega) \sim \omega ^{-\alpha}$.}
 
{ In this paper, we propose a 1D disordered multilayered structure in which 
 the random refractive index of one of its constituents follows a probability
  density function with a power-law tail with exponent $\alpha$ 
  (L\'evy-type distribution). The propagation of an obliquely incident electromagnetic wave through 
the structure is studied using the transfer matrix method. The effects of 
variation of the exponent $\alpha$ on the localization length and Anderson 
localization are investigated in different frequency ranges. It is shown 
that the localization length decreases with decreasing $\alpha$ for both $1<\alpha<2$ 
and $0<\alpha<1$ at intermediate and large frequencies. Furthermore, at small 
frequencies the localization length depends on values of $\alpha$. In this 
structure, the localization length can be also affected by the incident angle.}

The paper is organized as follows: in Sec. II we introduce
the notation. 
The numerical results are discussed in Sec. III. Finally, we draw our conclusions in Sec. IV.

\section{Definitions and settings}

\begin{figure}[t]
\begin{center}
\includegraphics[width=9cm,clip]{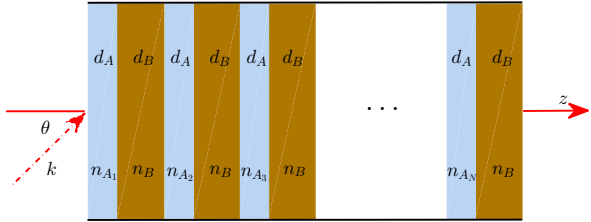}
\end{center}
\caption{(Color online)Sketch of a heterostructure of length $L$
with $N$ layers with the fixed thickness $d_{A(B)}$.
The refractive index $n_B$ is fixed and the distribution of 
the $n_A$s follows a L\'evy-type distribution.
}
\label{Figure:1}
\end{figure}

The disordered multilayered structure that we shall to study composed of 
an alternating sequence of
layers of $A$ and $B$ having thickness of $d_A$ an $d_B$. Fig. (\ref{Figure:1})
 displays a scheme of the proposed structure. 
 It is
assumed that the layers $B$ have the same refractive index $n_B$
 and the same width of $d_B$, while layers $A$
have the same thickness of $d_A$ and random refractive indices 
$n_{Aj}$($j=1,2,\dots,N$).
The number of layers in the structure is
taken to be $2N$ and the $z$-axis is directed across the layers. 

Here we consider $n_{Aj} \propto \sqrt{\mid \delta_j \mid}$, where $\delta_j$ are independent identically distributed  random variables 
with standard symmetric $\alpha$-stable distribution. 
The procedure of computer simulating realizations of the random variables $\delta_j$
is the following \cite{chambers1976method,weron1996chambers}:

\begin{eqnarray}\label{random_var_levy}
\delta_j = \frac{\sin \left(\alpha V \right)}{\left[ \cos\left( V\right)\right]^{1/\alpha}}
\left( \frac{\cos\left(V-\alpha V\right)}{W}\right)^{(1-\alpha)/\alpha}
\end{eqnarray}
where random variable $W$ has exponential distribution 
with mean $1$ and $V$ is uniformly distributed on $\left(-\pi/2,\pi/2\right)$.
The algorithm introduced in Eq. (\ref{random_var_levy}) allows us to generate a sequence of random numbers 
with $\alpha$-stable distribution for the whole range 
of parameter $0<\alpha \leqslant 2$. The main feature of a $\alpha$-stable L\'evy density distribution
$P(\delta)$ is the power-law, decay of its tail which behaves as: 
$P(\delta) \sim 1/\delta^{1+\alpha}$.

We choose random numbers $\delta_j$ which their absolute values are in the
range $5<\mid \delta_j \mid <100$. The refractive index of layer $A$
 is taken to be $n_{Aj} = \sqrt{\mid \delta_j \mid /5}$.

The transfer matrix formalism is used to compute the localization 
length of the structure. We
consider a monochromatic electromagnetic wave obliquely incident from left 
into the random
structure. In Fig. (\ref{Figure:1}), $k$ and $\theta$ denote the wave 
vector and incident angle, respectively. The wave vector $k$
is taken to be in $xz$-plane. The electric and magnetic fields at incident
 and exit ends of the structure
can be related by the product of transfer matrix of different layers 
included in the heterostructure as:
\begin{eqnarray}\label{} 
\begin{pmatrix}
  E_0 \\
  H_0 
 \end{pmatrix}= M_1 M_2 M_3 \dots M_{2N} 
\begin{pmatrix}
  E^{2N+1} \\
  H^{2N+1}
 \end{pmatrix} =
 M 
 \begin{pmatrix}
  E^{2N+1} \\
  H^{2N+1}
 \end{pmatrix},
\end{eqnarray}
where $M$ is the total transfer matrix of the system and
 $M_j$($j=1,2,\dots,2N$) is the transfer matrix of the $j^{\textrm{th}}$
dielectric layer which is defined as:
\begin{eqnarray}\label{} 
M_j = 
\begin{pmatrix}
  \cos \left( k_{zj}d_j \right) & -\frac{i}{\eta_j} \sin \left( k_{zj}d_j \right)\\
  -i\sin \left( k_{zj}d_j \right) & \cos \left( k_{zj}d_j \right)
   
 \end{pmatrix},
\end{eqnarray}
here $\eta^2_j =(\epsilon_0/\mu_0) \epsilon_j /\cos^2 \theta_j$ for TM case
 and $\eta^2_j =(\epsilon_0/\mu_0) \epsilon_j \cos^2 \theta_j$ for TE case, 
 $k_{zj} =k_j \cos \theta_j$ denotes
the $z$ component of the wave vector in dielectric layers, 
and $d_j$ is the thickness of different dielectric
layers. For a plane wave strikes from left into the disordered structure,
 the transmission coefficient
$t(\omega)$ is expressed in terms of the matrix element of $M$ as follows:
\begin{eqnarray}\label{•}
t(\omega) = \frac{2\eta_0}{M_{11}\eta_0 +M_{12}\eta_0 \eta_{N+1}+M_{21}+M_{22}\eta_{N+1}}.
\end{eqnarray}
The corresponding transmittance of the structure at frequency $\omega$ is:
\begin{eqnarray}\label{•}
T(\omega) = \frac{\eta_{N+1}}{\eta_0} \mid t(\omega) \mid ^2.
\end{eqnarray}
To study the localization behavior in the $1$D random system, 
it is required to evaluate the
localization length. Since the transmittance in the localized regime 
exponentially decays with the
system length $L$, the localization length $\xi$
can be numerically calculated as
\begin{eqnarray}\label{•}
\xi (\omega) = -\lim _{L\rightarrow \infty} \frac{2L}{\ln \left(T(\omega,L)\right)}.
\end{eqnarray}
For a sufficiently long random-layered system, $\xi$ obtained from the 
above equation is a nonrandom
number due to self-averaging. However, for a system with a finite size, 
the localization length can be
obtained by ensemble averaging of the transmittance $T$ over many 
realizations. This means that we
introduce the localization length of a finite random configuration as
\begin{eqnarray}\label{•}
\xi (\omega) = - \frac{2L}{\langle\ln \left(T(\omega,L)\right)\rangle}.
\end{eqnarray}
where $\langle  \dots \rangle$ stands for the ensemble averaging.
The values of those parameters used in the following calculations are 
$d_A=20\textrm{mm}$, $d_B=40\textrm{mm}$, $n_B=1$ and
$N=500$. In the next section, we will discuss our numerical findings.

\section{Results and discussion}
\begin{figure}[t]
\begin{center}
\includegraphics[width=9cm,clip]{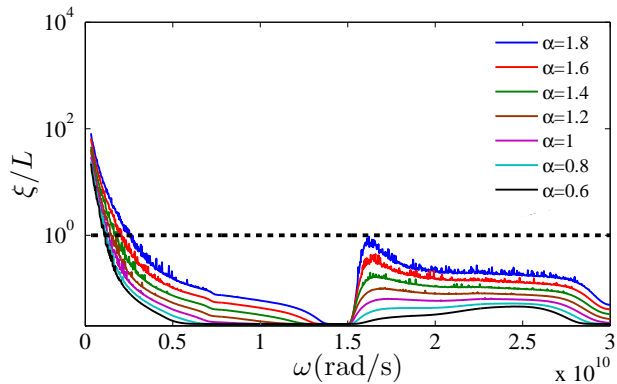}
\end{center}
\caption{(Color online) Localization lengths in units of the system size
 for normal incidence in the frequency range 
 $3\times10^8 \textrm{(rad/s)}<\omega<3\times10^{10} \textrm{(rad/s)}$ 
 for different $\alpha$ values.}
\label{Figure:2}
\end{figure}

We first study the case at which the plane wave is normally incident 
into the random structure. It is
assumed the frequency of the incident wave is in the range 
$3\times10^8 \textrm{(rad/s)}<\omega<3\times10^{10} \textrm{(rad/s)}$. In order to
understand how localization length can be affected by the variation of 
$\alpha$ values, in Fig. (\ref{Figure:2}), we plot the
localization length in units of the system length 
(normalized localization length) as a
function of frequency $\omega$ for different values of $\alpha$. 
To obtain localization length, $2\times10^4$ different
random realizations with the same $\alpha$ values and the same number of 
layers are considered. 
As shown
in Fig. (\ref{Figure:2}), the localization length decreases with 
decreasing the exponent $\alpha$ at some frequency ranges. For
frequencies at which normalized localization length of the wave is 
lower than one, the system is in the
localized regime. When $\alpha$ decreases from $1.8$ to $0.6$, 
the minimum frequency at which localization
occurs shifts toward lower frequencies. 
{Our numerical results confirm the self-averaging of the localization length. 
That is the localization length obtained from Eq. (7) does not significantly 
differ from the localization length of a single realization with large 
number of layers. Furthermore, increasing the number of random realizations 
from $2\times 10^4$ does not lead to any considerable change in localization 
length values.}  

\begin{figure}[t]
\begin{center}
\includegraphics[width=9cm,clip]{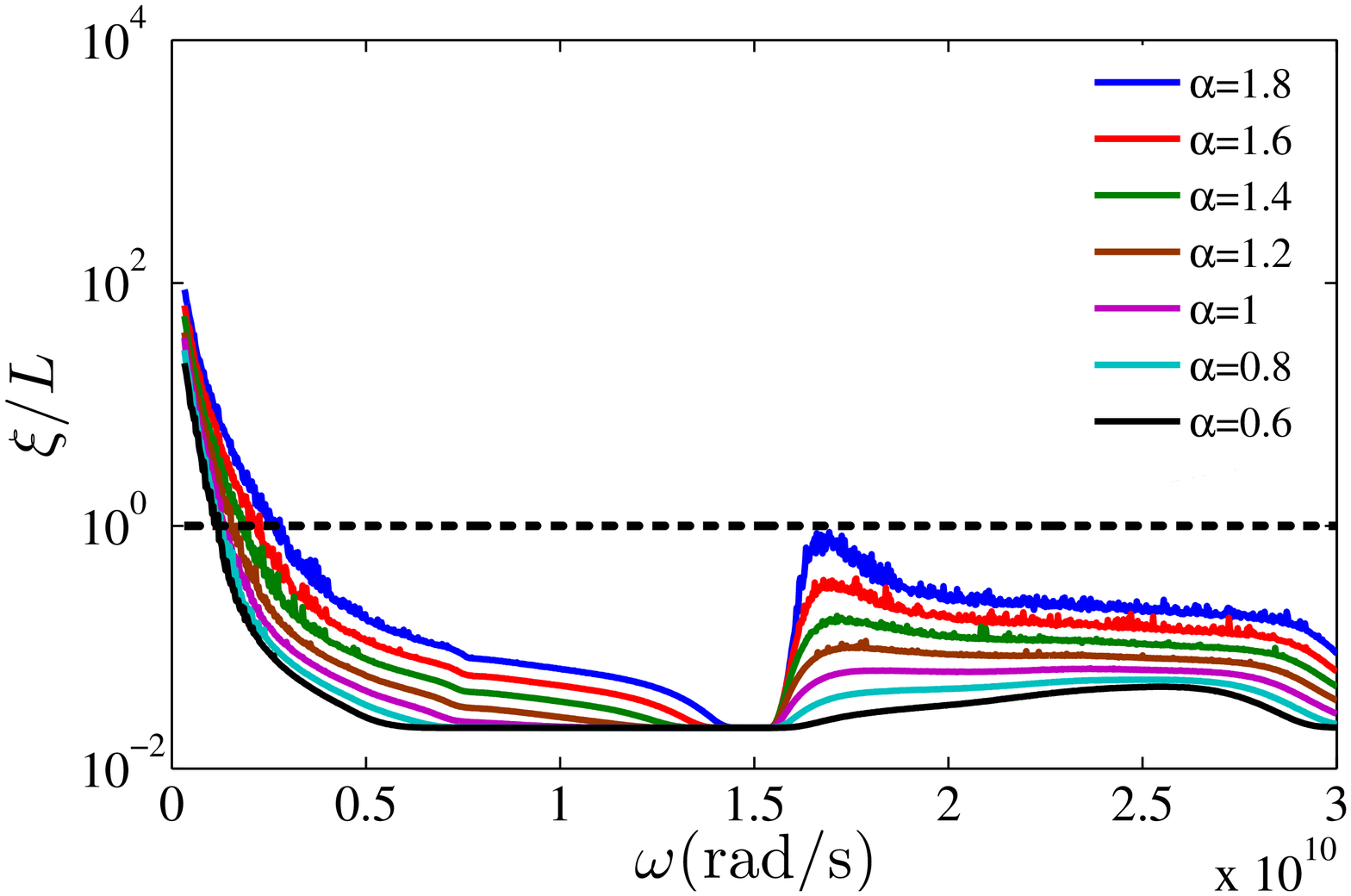}
\includegraphics[width=9cm,clip]{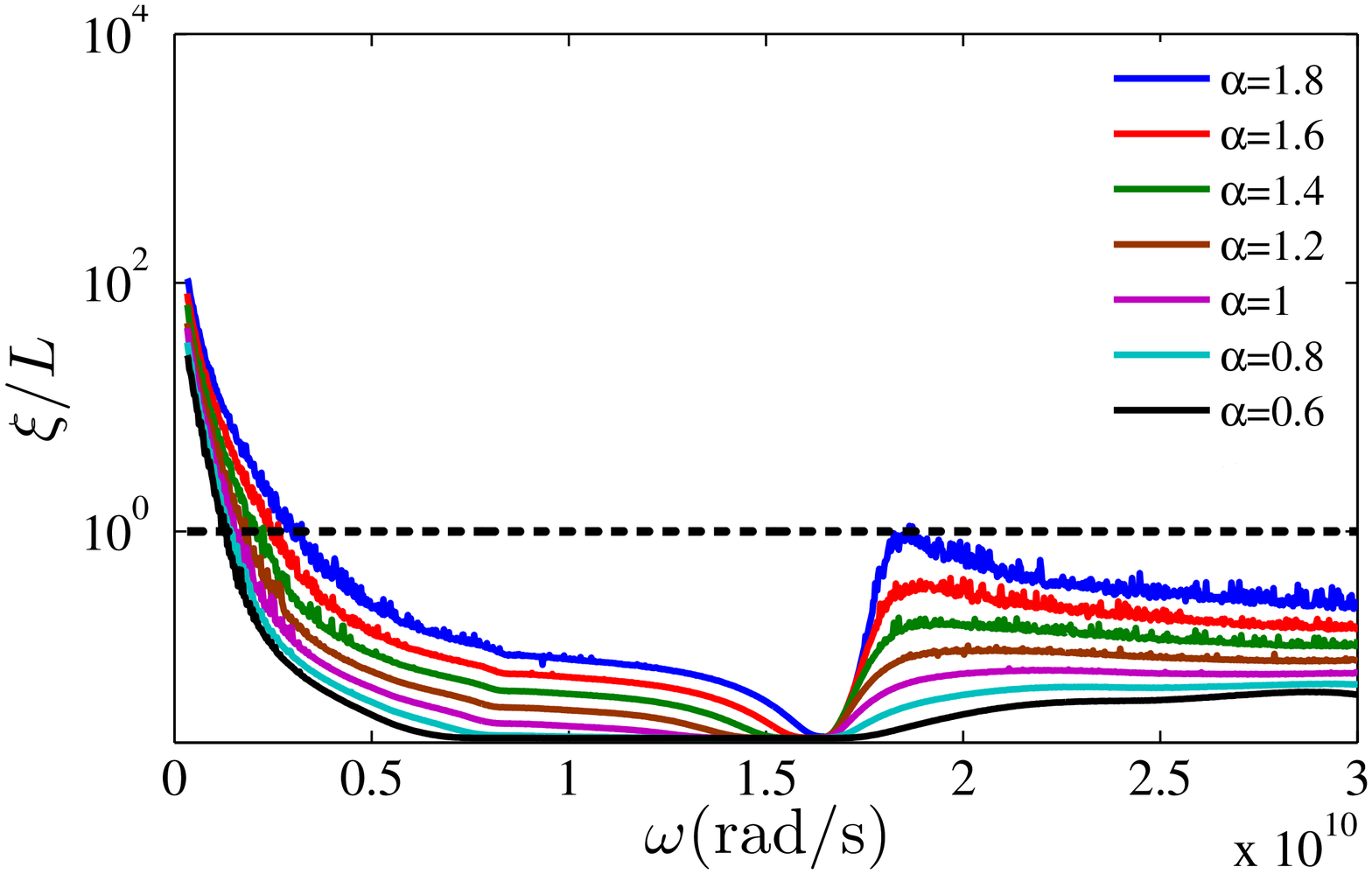}
\end{center}
\caption{(Color online) Localization lengths in units of the system size
 for oblique incidence $\theta=15^{\circ}$ (top) and $\theta=30^{\circ}$ (bottom) 
 in the frequency range 
 $3\times10^8 \textrm{(rad/s)}<\omega<3\times10^{10} \textrm{(rad/s)}$ 
 for different $\alpha$ values.}
\label{Figure:3}
\end{figure}

\begin{figure}[t]
\begin{center}
\includegraphics[width=9cm,clip]{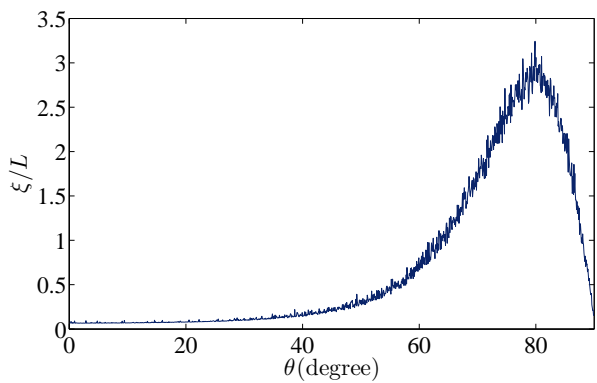}
\end{center}
\caption{{(Color online)Localization lengths in units of the system size
 versus incident angle $\theta$ for $\alpha = 0.6$ and 
 $\omega = 2\times10^9 \textrm{(rad/s)}$. 
}}
\label{Figure:5}
\end{figure}

To determine whether or not this dependence of localization length 
on the $\alpha$ values is observed in
the case of oblique incidence, we display the normalized localization 
length versus frequency for
$\theta=15^\circ$ and $\theta=30^\circ$ in Fig. (\ref{Figure:3}).
The corresponding frequency range is the same as that
in Fig. (\ref{Figure:2}). It is clearly seen in Fig. (\ref{Figure:3}) 
that for oblique incidence the localization length decreases
when the exponent $\alpha$ decreases from $1.8$ to $0.6$. 
To better understand the effect of incident angle on the localization
length, in Fig. (\ref{Figure:5}) we show the normalized localization 
length versus $\alpha$ for different incident angles $\theta=0$, $15^\circ$ and $30^\circ$
 with the same $\alpha=1.8$.
\begin{figure}[t]
\begin{center}
\includegraphics[width=9cm,clip]{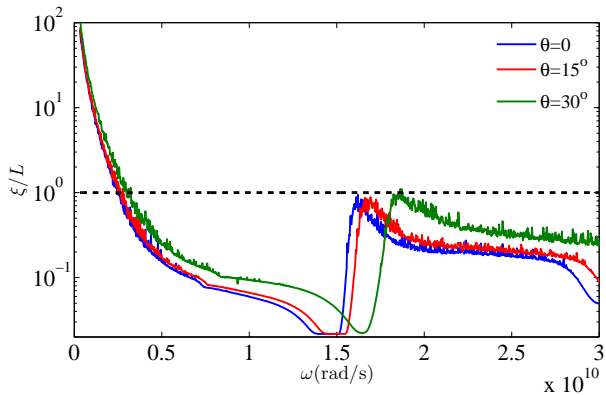}
\end{center}
\caption{(Color online)Localization lengths in units of the system size
 for different incident angles $\theta=0$, $15^\circ$ and $30^\circ$  in the frequency range 
 $3\times10^8 \textrm{(rad/s)}<\omega<3\times10^{10} \textrm{(rad/s)}$ 
 with the same $\alpha$ values.}
\label{Figure:4}
\end{figure}

One can see that the localization length increases with incident angle
at some frequency regions. Furthermore, the dip in localization length 
moves to higher frequencies
with increasing incident angle. Therefore, our calculated results 
indicate that the localization length
depends on the incident angle as well as the exponent $\alpha$.

{
It is well known that for TM polarized waves propagating in a 1D random 
structure the localization length takes a large maximum value at some 
critical angles which are called generalized Brewster angles} \cite{brewster1}. {It has 
been demonstrated that generalized Brewster angle increases from $0^\circ$ to $90^\circ$ 
with increasing the disorder-averaged refractive index }\cite{brewster1}. {The localization 
length for a weak disorder diverges in the vicinity oh $\theta=45^\circ$ when the average of refractive index is equal to 1 }
\cite{brewster1,brewster2}. {This phenomenon is known as Brewster anomalies and the corresponding 
angle is called the Brewster angle} \cite{brewster1,brewster2}. {In Fig.} (\ref{Figure:5}), {we display the 
localization length versus incident angle for $\alpha=0.6$ and $\omega=2\times10^9 \textrm{(rad/s)}$. 
One can see that the generalized Brewster angle is about $80^\circ$ at which the 
localization length is significantly enhanced over its value at the normal 
incidence ($\theta=0^\circ$). Moreover, for incident angle in the range $68^\circ<\theta<88^\circ$, the 
system is in the extended regime, while at other incident angle the system 
is in the localized regime. The effects of $\alpha$ variation on the generalized 
Brewster angle and Brewster anomalies are under investigation and their 
results will be reported in the near future.}

\begin{figure}[t]
\begin{center}
\includegraphics[width=9cm,clip]{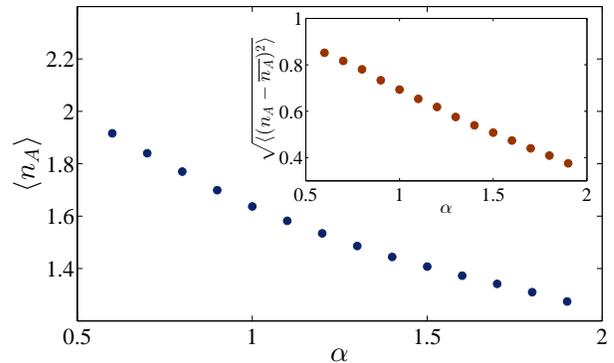}
\end{center}
\caption{(Color online) Mean and variance (inset) values of refractive index $n_A$ 
as a function of the exponent $\alpha$. 
}
\label{Figure:6}
\end{figure}

To understand the physical reason for the behavior of localization 
length with $\alpha$, in Fig. (\ref{Figure:6}), we plot
the mean value of refractive index versus $\alpha$. As shown in 
Fig. (\ref{Figure:6}), the average value of refractive index
increases with increasing $\alpha$. As a result, increases of $\alpha$ leads 
to the decreasing the refractive
index contrast between the layers of disordered system. 
This effect causes the scattering strength to
decrease. Hence, we expect that the localization length decreases with 
increasing $\alpha$. In addition, the Fig. (\ref{Figure:6}) 
represents the variance of refractive index versus $\alpha$.
 One can see that the variance of refractive index increases with
  decreasing $\alpha$. Therefore, the randomness strength increases with 
  decreasing $\alpha$.
This effect also results in the enhancement of localization with 
decreasing $\alpha$.

\begin{figure}[t]
\begin{center}
\includegraphics[width=9cm,clip]{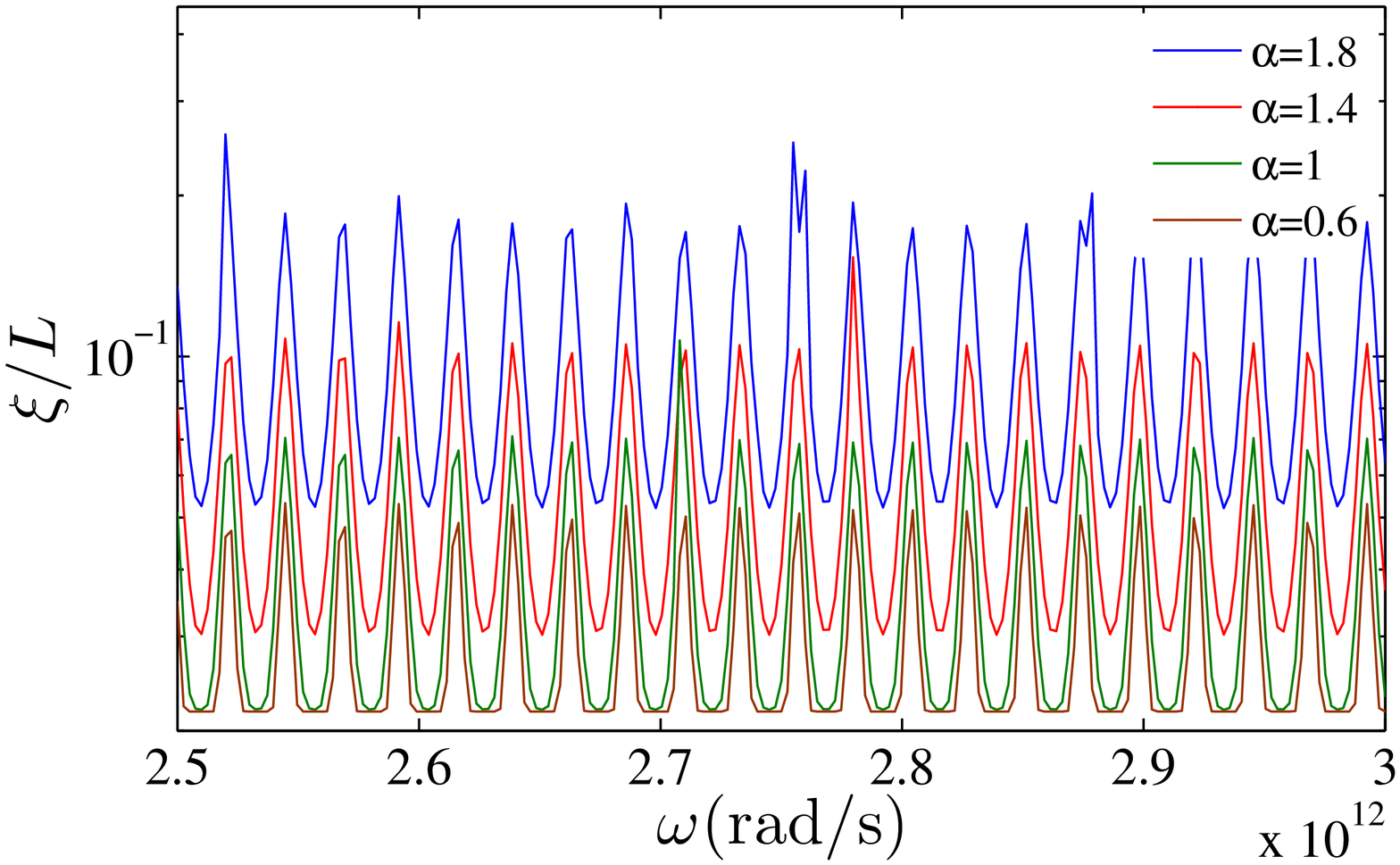}
\end{center}
\caption{(Color online) Localization lengths in units of the system size
 for normal incidence in the frequency range 
 $2.5\times10^{12} \textrm{(rad/s)}<\omega<3\times3^{12} \textrm{(rad/s)}$ 
 for different $\alpha$ values.}
\label{Figure:7}
\end{figure}

Now, we shall to study how heavy-tail distribution of random refractive index
$n_A$ affects the normalized localization length in the frequency
range $2.5\times 10^{12} \textrm{(rad/s)}<\omega<3\times 10^{12} \textrm{(rad/s)}$. Fig. (\ref{Figure:7}) shows 
the corresponding results for different values of
$\alpha=1.8$, $1.4$, $1.0$ and $0.6$. 

It is clearly seen that the normalized localization length shows an 
oscillatory
behavior in this frequency range for different $\alpha$ values. 
Moreover, for all frequencies in this frequency
range and for all $\alpha$ values, we have a localized mode whose 
localization length decreases with
decreasing $\alpha$. As a result, decreasing $\alpha$ value improves 
the localization. This effect is attributed to the
enhancement of mean value and variance of refractive index with decreasing $\alpha$.

\begin{figure}[t]
\begin{center}
\includegraphics[width=9cm,clip]{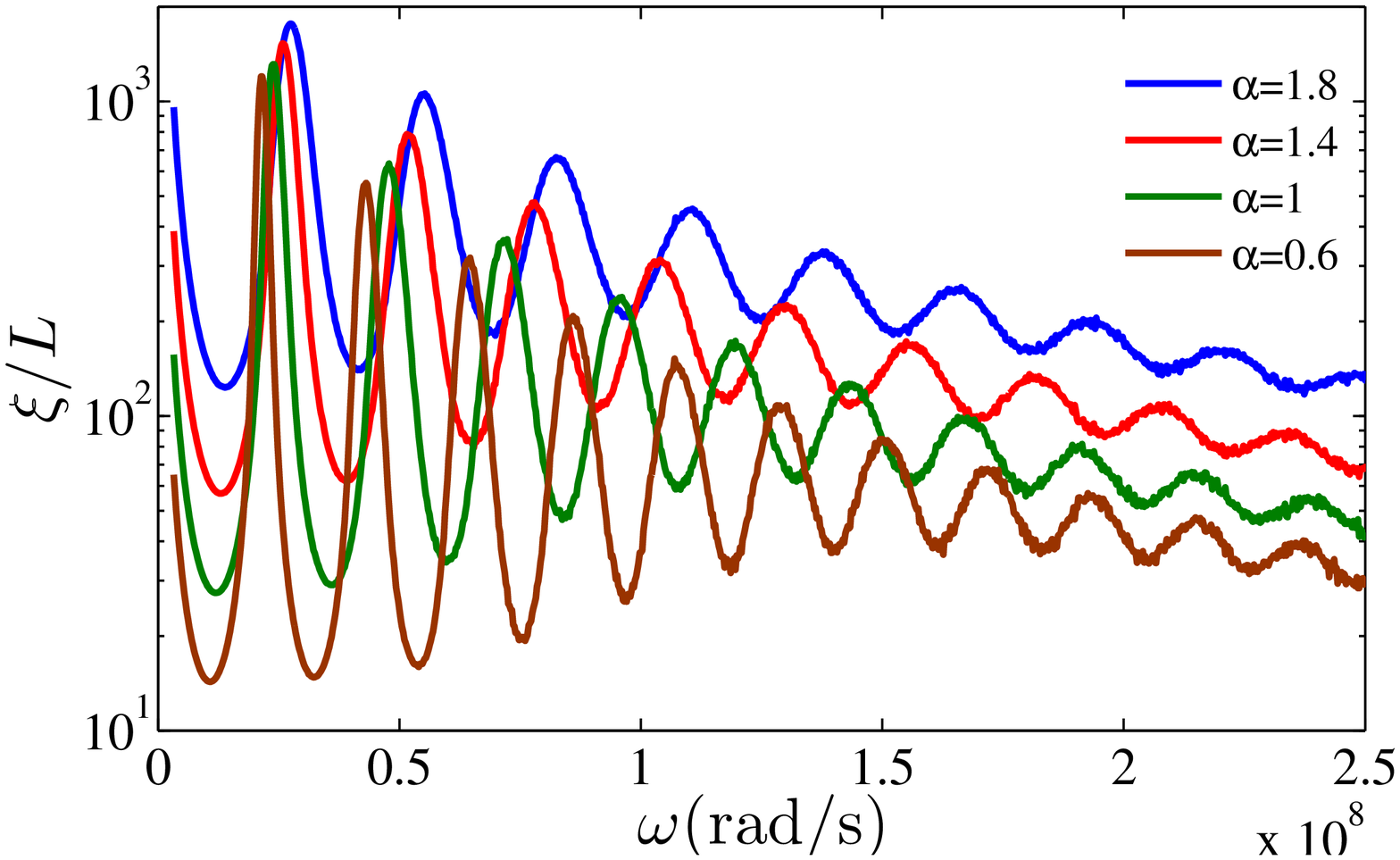}
\end{center}
\caption{(Color online) Localization lengths in units of the system size
 for normal incidence in the frequency range 
 $3\times10^6 \textrm{(rad/s)}<\omega<2.5\times10^{8} \textrm{(rad/s)}$ 
 for different $\alpha$ values.}
\label{Figure:8}
\end{figure}

Next, we consider the effect of $\alpha$ variation on the normalized 
localization length for longer
wavelengths. The normalized localization lengths versus $\omega$ in frequency
 range $3\times10^6 \textrm{(rad/s)}<\omega<2.5\times10^8 \textrm{(rad/s)}$
  are displayed in Fig. (\ref{Figure:8}) for different values of 
  $\alpha=1.8$, $1.4$, $1.0$ and $0.6$. 
  
  As shown in
this figure, the localization length indicates an oscillatory behavior 
where their peak values increases
with lowering the frequency for all $\alpha$ values. For all wavelengths, 
the system is in the extended
regime. With decreasing the $\alpha$ values, the peaks of normalized 
localization length shift to lower
frequencies, while the peak values decreases.

Consequently, our results demonstrate that if the distribution of the 
random layer in a $1$D disordered
system follows the $\alpha$-stable L\'evy distribution, the localization 
length can be manipulated with the exponent $\alpha$.
 Due to the random nature of Anderson localization, the control of
this phenomenon in a regular manner is of great importance and has 
potential applications.

\section{Conclusion}
We have studied the localization of an electromagnetic wave normally and 
obliquely incident into a $1$D disordered structure where the refractive 
index of one its constituents is fixed while that of the
other constituents is a random number drawn from a L\'evy-type 
distribution with exponent $\alpha$. It has
been demonstrated that for normal incidence the localization behavior in 
this structure can be
manipulated in a regular manner by changing the value of $\alpha$. 
When $\alpha$ decreases, the localization
length of the waves decreases in the frequency range $3\times10^8 \textrm{(rad/s)}
<\omega<1.9\times10^{10} \textrm{(rad/s)}$ decreases. This
effect is due to the increase of mean and variance of refractive index 
with decrease of $\alpha$. Moreover,
the minimum frequency at which localization appears shifts to lower 
frequencies with decreasing $\alpha$.
Localization length shows the same trend with variation of $\alpha$ 
for oblique incidence. It has been also
investigated how the localization length can be affected by $\alpha$ 
variation in the lower and higher
frequency ranges. In the frequency range $2.5\times10^{12} \textrm{(rad/s)}
<\omega<3\times 10^{12} \textrm{(rad/s)}$, the system is in the localized
regime and the localization length indicate an oscillatory behavior with 
approximately fixed
amplitudes. Decrease of $\alpha$ in this frequency range also gives rise 
to the enhancement of localization.
In the frequency range $3\times10^6 \textrm{(rad/s)}<\omega<2.5
\times10^8 \textrm{(rad/s)}$. it is found the system is in the 
extended regime
and the localization length exhibits an oscillatory behavior with 
increasing amplitude whose value
decreases with decreasing $\alpha$. Furthermore, reduction of $\alpha$
 causes the peaks of localization length to
shift toward lower frequencies. 
Consequently, in disordered media, employing the $\alpha$-stable L\'evy
distribution provides a way to easily control the localization phenomenon.

%\bibliography{qie}
\end{document}